\documentclass[aps,twocolumn,10pt,nofootinbib]{revtex4}% twocolumn,article, book, report,nofootinbib
\usepackage{graphicx,amssymb,amsmath}%amsart,hyperref,href

\newcommand{\braket}[1]{\left<#1\right>}
\newcommand{\para}[1]{\left(#1\right)}
\newcommand{\abs}[1]{\left|#1\right|}

\begin{document}

\title{Isotope effect on $T_c$ and the superfluid density of high-temperature superconductors}
\author{Abolhassan Vaezi}
\affiliation{Department of Physics,
Massachusetts Institute of Technology,
%77 Massachusetts Avenue,
Cambridge, MA 02139, USA}
\email{vaezi@mit.edu}

\date{\today}

\begin{abstract}
In this paper, I study the oxygen isotope effect (OIE) in cuprates. I introduce a simple model that can explain experiments both qualitatively and quantitatively. In this theory, isotope substitution only affects the superfluid density, but not the pseudo-gap. Within the spin-charge separation picture, I argue that the spinon-phonon interaction is in the adiabatic limit, and therefore within the Migdal-Eliashberg theory, there is no isotope effect in the spinon mass renormalization. On the other hand, I show that the holon-phonon interaction is in the non-adiabatic limit. Therefore, the small polaron picture is applicable and there is a large mass enhancement in an isotope-dependent way. Our theory explains why upon $^{16}$O/$^{18}$O substitution, the superconducting transition temperature changes only in underdoped cuprates, while there is no considerable OIE at the optimal doped as well as the overdoped cuprates. Additionally, in contrast to the conventional superconductors, we obtain OIE on the superfluid density for whole superconducting region in agreement with the experimental observations.
\end{abstract}
%\langle \rangle
\maketitle
%\chapter{just for report.cls}
%\part{can be used in RevTEX.cls as well}
%\section{t-J Model+Phonon mediated holon-holon pairing}

\section{Introduction}

Finding the underlying mechanism of the high temperature superconductivity in cuprates$~$\cite{Bednorz_Mueller_1986} is one of the most challenging and outstanding problems in theoretical physics. Observation of strong isotope effect on both the transition temperature and the superfluid density in cuprates$~$\cite{Zimmerman_1,Khasanov_3_2008,Raffa_1,Zhao_1}, indicates the importance of the electron-phonon interaction in understanding the physics of high Tc$~$\cite{Schneider_Keller_1992_a,Batlogg_1}. Experiments show strong OIE on $T_c$ only in underdoped cuprates (see Fig. 1), while there is no considerable OIE in overdoped cuprates. On the other hand OIE on the London penetration depth ($\lambda_{ab}\para{0}$) (in-plane penetration depth) has been reported for both sides $~$\cite{Hofer_2000_a,Khasanov_1_2004,Tallon_1,Khasanov_4_2003}. In both cases the isotope exponent decreases as we approach the optimal doping from the underdoped side. There is also an unusual correlation between isotope effect on $T_c$ and $\lambda_{ab}\para{0}$ \cite{Khasanov_2_2006}(See Fig. 2). It is impossible to explain such effects using BCS theory$~$\cite{BCS} or its extensions such as Migdal-Eliashberg$~$\cite{Marsiglio_1} theory. The reason is that within BCS theory which is based on the adiabatic electron-phonon approximation$~$\cite{Khasanov_3_2008}, the isotope exponent of $T_c$ which is defined as $\alpha=-\frac{M}{T_c}\frac{\Delta T_c}{\Delta M}$, is around 1/2$~$\cite{BCS}. On the other hand, the electron effective mass is $m^*=m\para{1+\lambda}$$~$\cite{Maksym_PA_Lee_2010_a}, where $\lambda$ is the dimensionless phonon mediated attraction coupling, and is isotope independent. Therefore, theories based on the adiabatic approximation, predict absence of the isotope effect on the superfluid density $(n_s/m^*)$ which is inconsistent with experiments. Therefore, we have to look for other theories of superconductors. In this paper we start from Anderson theory of high Tc$~$\cite{Anderson_1987Sci} which is based on the strong correlation physics and the spin fluctuation pairing mechanism. This theory is very successful in explaining many aspects of cuprates but it does not take phonons into account. It is very important to explain the oxygen isotope effect within this successful theory. Here we use this general framework and then by adding electron phonon interaction, it is shown that the observed unusual isotope effects on $T_c$ as well as $\lambda_{ab}\para{0}$ can be explained.

 We start from t-J-Holstein model as our model Hamiltonian. In our model for simplicity, electrons are only coupled to a single Einstein phonon mode ($\omega_{_{E}}$). Then we use the spin-charge separation picture and we implement it by the slave boson method. Our idea is as follows: in underdoped cuprates, the superconducting transition temperature is controlled by the superfluid density $(n_{s}/m^*)$, which is mostly determined by that of holons, in our theory. On the other hand in overdoped cuprates, $T_c$ is controlled by the pseudo-gap which is equal to pairing order parameter of spinons $\Delta_s$ in our theory. Electron-phonon interaction affects the superfluid by holon mass renormalization in an isotope-dependent way, while it does not affect the pseudo-gap much. In our theory, we argue that holons are strongly coupled to phonons and therefore, we are in the non-adiabatic limit, while spinons are weakly coupled to phonons and we should use the adiabatic limit calculation (Migdal-Eliashberg theory) for them. Therefore, $E_{b}<\omega_{_{E}}<E_{s}$, where $E_{b}$ and $E_{s}$ are typical energies of holons and spinons respectively. It is also shown that in the presence of spinon pairing (pseudo-gap), the holon-phonon coupling constant,$\gamma$, renormalizes to $\Delta_{s}\gamma$. These facts together can explain the observed OIE on $T_c$ as well as on $\lambda^{-2}\para{0}\propto \frac{n_{s}}{m_{*}}$.

\begin{figure}
  \includegraphics[width=210pt]{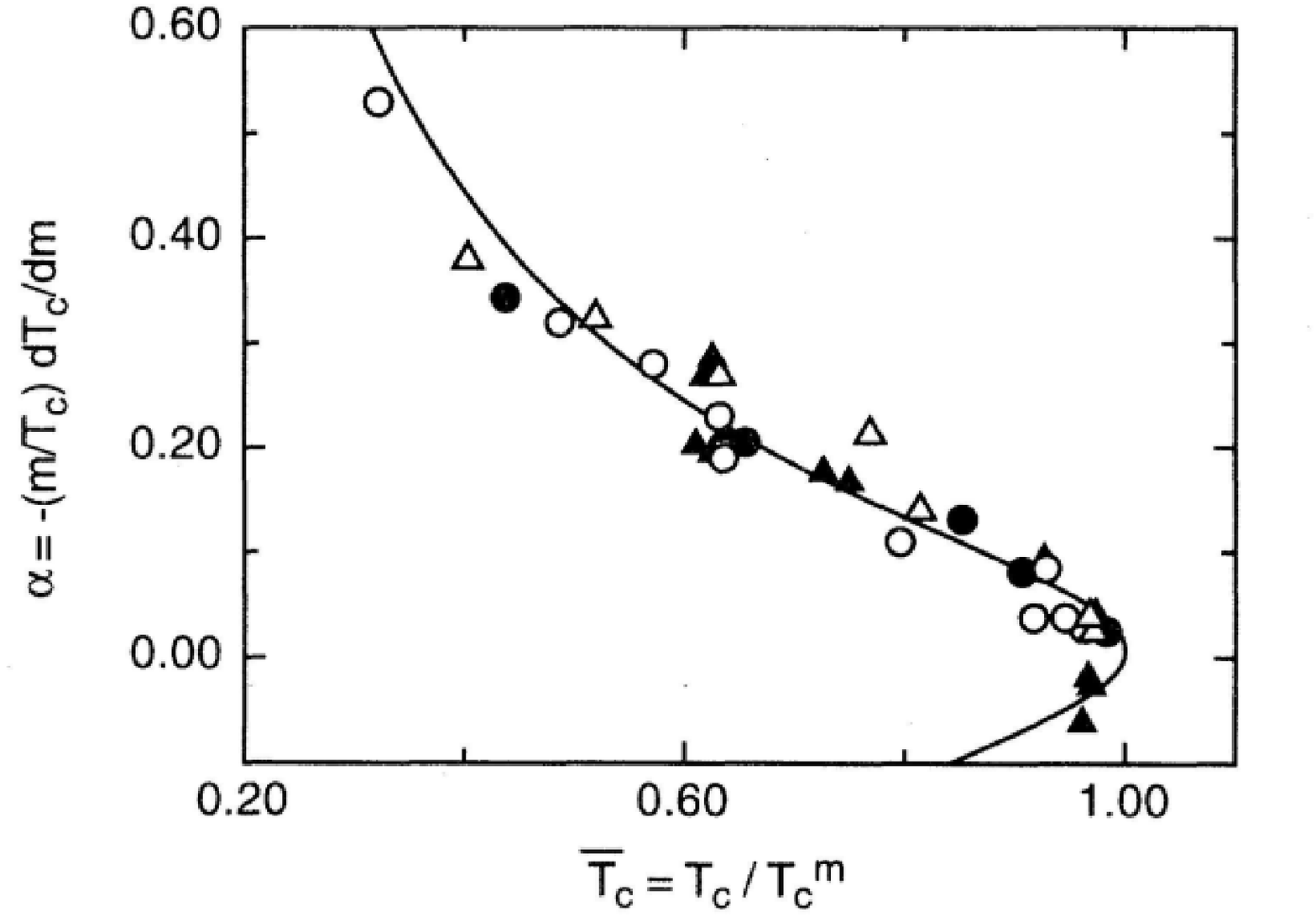}\\
  \caption{ Measured oxygen isotope coefficient of $T_c$ $(\alpha)$ for doped YBCO as a function of the reduced transition temperature ($\bar{T_c}=T_c/T_{c,max}$) from different samples. After[3]}\label{fig}
\end{figure}

\begin{figure}
  \includegraphics[width=180pt]{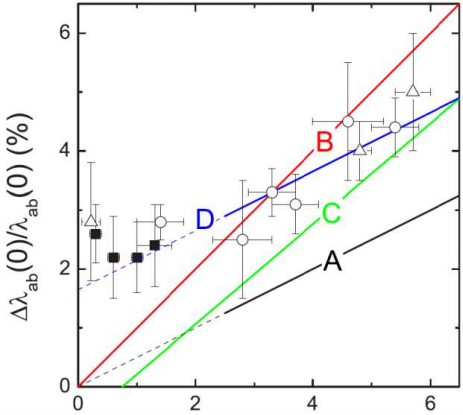}\\
  \caption{ A color plot of the OIE shift $\Delta \lambda_{ab}\para{0}/\lambda_{ab}\para{0}$
versus the OIE shift $-\Delta T_c /T_c$ for Y$_{1-x}$Pr$_x$Ba$_2$Cu$_3$O$_{7-\delta}$, YBa$_2$Cu$_4$O$_8$,
and La$_{2-x}$SrxCuO$_4$. Squares are the $\mu$SR data obtained in the
present study. Circles are bulk $\mu$SR data for Y$_{1-x}$Pr$_x$Ba$_2$Cu$_3$O$_{7-\delta}$
(Refs. 4 and 12 ) and LE $\mu$SR data for optimally doped
Y$_{1-x}$Pr$_x$Ba$_2$Cu$_3$O$_{7-\delta}$ (Ref. 10 ). Triangles are torque magnetization and
Meissner fraction data for La2-xSrxCuO4 (Refs. 6 and 9). Different lines are discussed in Ref. 13}\label{fig}
\end{figure}

\section{Method}

Let us start from the t-J\cite{Lee_Nagaosa_Wen_2006a,PA_Lee_1992_1} model which is defined as:

\begin{eqnarray}
  H_{t-J}=-t\sum_{\left<i,j\right>, \sigma }P_{G}c_{\sigma,i}^\dag c_{\sigma,j}P_{G}+J\sum_{i,j}\hat{S}_{i}.\hat{S}_{j}
\end{eqnarray}

where $P_{G}$ is the Gutzwiller projection operator which removes doubly occupied states. Within the slave boson formalism, electrons can be decomposed as $c_{i,\sigma}^\dag = f_{i,\sigma}^\dag h_{i} $ along with the physical constraint on each site:  $h_{i}^\dag h_{i}+\sum_{\sigma}f_{i,\sigma}^\dag f_{i,\sigma}=1$ which implements the Gutzwiller projection. Now if we decouple spinons (spin sector) from holons (charge sector), we can write two following effective Hamiltonians for each sector:

\begin{eqnarray}
  &&H_{h}= -\sum_{<i,j>}t\chi_{s} h_{i}^\dag h_{j}-\sum_{i}\mu_{h}h_{i}^\dag h_{i}
\end{eqnarray}
\begin{eqnarray}
  H_{s}=&&-\sum_{<i,j>,\sigma}\para{t\chi_{h}+\tilde{J}\chi_{s}}f_{i,\sigma}^\dag f_{j,\sigma} -\sum_{i,\sigma}\mu_{s}f_{i,\sigma}^\dag f_{i,\sigma}\cr &&-\sum_{<i,j>}~2\tilde{J}\Delta_{s}\para{i,j}\para{f_{i,\uparrow}^\dag f_{j,\downarrow}^\dag-f_{i,\downarrow}^\dag f_{j,\uparrow}^\dag}  +h.c.~~~~
\end{eqnarray}

where the following notations have been used:

\begin{eqnarray}
&&  \chi_h=\braket{h_{i+\vec{\delta}}^\dag h_{i}}\simeq p\\
&&  \chi_f=\braket{\sum_{\sigma}f_{i+\vec{\delta},\sigma}^\dag f_{i,\sigma}}\\
&&  \Delta_{s}\para{i,j}=\frac{1}{2}\braket{f_{i,\uparrow}^\dag f_{j,\downarrow}^\dag-f_{i,\downarrow}^\dag f_{j,\uparrow}^\dag}
\end{eqnarray}

Form the t-J model one may expect that $\tilde{J}=J/4$ but in literature it has been discussed that the best choice is $\tilde{J}=3J/8$\cite{Lee_Nagaosa_Wen_2006a}. This model has been studied by many authors in the past three decades. It is well known that this model can lead to the d-wave pairing of spinons\cite{Lee_Nagaosa_Wen_2006a}, {\em i.e.} $\Delta_{s}\para{\pm \hat{x}}=\Delta_{s}$ and $\Delta_{s}\para{\pm \hat{y}}=-\Delta_{s}$.

Now let us consider the electron-phonon interaction. To do so we add the following Hamiltonian to the t-J model.

\begin{eqnarray}
  H'=\sum_{q} \omega_{_{E}} a_{q}^\dag a_{q} +\sum_{k,q}\gamma_{k,q}\para{a_{q}+a_{-q}^\dag}\tilde{c}_{k+q,\sigma}^\dag\tilde{c}_{k,\sigma}
\end{eqnarray}

where $\tilde{c}_{k,\sigma}$ is the Fourier components of $P_{G}c_{i,\sigma}P_{G}=f_{i,\sigma} h_{i}^\dag$. If we treat $\tilde{c}_{k,\sigma}$ as a fermionic operator, then since the typical energy of electrons is around $J$ and is much larger than $\omega_{_{E}}$, so that we can apply the BCS theory, or the Migdal-Eliashberg theory as its extension. BCS approximation lead to the following pairing term:

\begin{eqnarray}
  -\sum_{k,k'}V_{k,k'}<\tilde{c}_{k',\uparrow}^\dag \tilde{c}_{-k',\downarrow}^\dag>\tilde{c}_{-k,\downarrow} \tilde{c}_{k,\uparrow}
\end{eqnarray}

Now if we translate everything to the slave boson language in real space, within mean-field approximation we can substitute $\tilde{c}_{i,\downarrow} \tilde{c}_{j,\uparrow}=f_{i,\downarrow} f_{j,\uparrow}h_{i}^\dag h_{j}^\dag$ by the following form:

\begin{eqnarray}
\Delta_{h}\para{i,j} f_{i,\downarrow} f_{j,\uparrow} + \Delta_{s}\para{i,j}h_{i}^\dag h_{j}^\dag-\Delta_{s}\para{i,j}\Delta_{h}\para{i,j}
\end{eqnarray}

where $\Delta_{h}\para{i,j}=\braket{h_{i}h_{j}}$. Now assuming a very short range interaction, we obtain the following effective interaction:

\begin{eqnarray}
&&  H'_{s-s}=-V \Delta_{h}^2 \sum_{<i,j>}\Delta_{s}\para{i,j}f_{i,\uparrow}^\dag f_{j,\downarrow}^\dag+h.c.\\
&&  H'_{h-h}=-V \Delta_{s}^2 \sum_{<i,j>}\Delta_{h} h_{i}^\dag h_{j}^\dag +h.c.
\end{eqnarray}

where $\Delta_{h}\sim p$ (doping). Let us assume that $V\Delta_{h}^2 \sim V p^2 \ll J$, so we can neglect this phonon mediated pairing term and therefore, the d-Wave nature of the spinons does not change.  From the above we see that the coupling constant of phonon mediated spinon spinon attraction is renormalized by $\Delta_{h}^2$ factor and that of holons by $\Delta_{s}^2$ due to strong correlation effects. It is easy to show that, $V\propto \gamma^2$, therefore, we can interpret these renormalization factors as the renormalization of the coupling constant of spinon-phonon interaction to $\Delta_{h}\gamma_{k,q}$ and that of holon-phonon interaction to $\Delta_{h}\gamma_{k,q}$. Therefore, we can substitute the electron phonon interaction term by sum the following two terms:

\begin{eqnarray}
  &&H_{s-ph}=\sum_{k,q}\Delta_{h}\gamma_{k,q}\para{a_{q}+a_{-q}^\dag}f_{k+q,\sigma}^\dag f_{k,\sigma}\\
  &&H_{h-ph}=\sum_{k,q}\Delta_{s}\gamma_{k,q}\para{a_{q}+a_{-q}^\dag}h_{k+q}^\dag h_{k}
\end{eqnarray}

Now let us consider spinon-phonon interaction. The typical energy of spinons is around $J$. $J$ is around 1500 Kelvin while the typical energy of optical phonons ($\omega_{_{E}}$) is a few hundred kelvin. Therefore, spinon-phonon interaction is in the adiabatic limit and therefore, we can apply Migdal-Eliashberg theory. In this regime the mass renormalizes as: $m^*=m\para{1+\lambda}$, where:

\begin{eqnarray}
  \lambda=2\int \frac{d\omega}{\omega}\alpha^2\para{\omega}F\para{\omega}
\end{eqnarray}

From our simple model it can be shown that $\lambda \propto \frac{1}{M\omega_{_{E}}^2}$, and since $\omega_{_{E}}\propto \frac{1}{\sqrt{M_{ion}}}$, $\lambda$ is isotope independent. Therefore, oxygen isotope substitution does not enhance the effective mass of spinons. This agrees with experiment where there is no isotope effect on Fermi velocity by Laser ARPES, while there is shift in kink energy$~$\cite{Iwasawa_1}.

On the other hand, holons are hard-core bosons. They usually condense at the bottom of their energy band. So their effective band-width is much smaller than their actual band-width. To have a better idea, for the moment let us assume that they are fermions. In that case their Fermi energy will be around $4\chi_{f}tp$, where $p$ is doping. Therefore, their effective bandwidth is at most of the order of $tp\chi_{f}$. Now if $\omega_{_{E}}$ is larger than the typical kinetic energy of holons, the electron phonon interaction is in the non-adiabatic limit. In summary if $tp\chi_{f}<\omega_{_{E}}<J$ then spinon-phonon interaction is in the adiabatic limit and holon-phonon interaction is in the non-adiabatic limit. For this limit it is easier to rewrite $H_{h}+H_{ph}+H_{h-ph}$ in the following way:

\begin{eqnarray}
H=&& -\sum_{<i,j>}t\chi_{s} h_{i}^\dag h_{j}-\sum_{i}\mu_{h}h_{i}^\dag h_{i} +\sum_{q} \omega_{_{E}} a_{q}^\dag a_{q}\cr
&&+\frac{1}{\sqrt{2NM\omega_{_{E}}}}\sum_{q,n,\sigma} \Delta_{s}\gamma_{n,q} h_{n}^\dag h_{n} \para{d_{-q}^\dag +d_{q}}
\end{eqnarray}

The above model has been extensively studied. In the non-adiabatic regime, we can use the results for a single polaron theory$~$\cite{Alexanrov_1,Alexanrov_2}. Therefore, we can do the powerful Lang-Firsov transformation$~$\cite{Hohenadler_1}, {\em i.e.} $H\rightarrow \tilde{H}=e^{-S}He^{S}$ where:  $S=\frac{1}{{\omega_{_E}}\sqrt{2NM\omega_{_{E}}}}\sum_{q,n} \gamma_{n,q}\Delta_{s} h_{n}^\dag h_{n} \para{d_{-q}^\dag -d_{q}}$. As an approximation let us assume that $<\gamma_{n,q}>=\gamma_{0}\frac{exp\para{ik.\vec{R}_n}}{\sqrt{N}}$. Therefore, we finally have:

\begin{eqnarray}
  &&\tilde{H} \simeq  -\sum_{\vec{k}}\para{4t\tilde{\chi}_{s}\para{\cos{k_x}+\cos{k_y}}+\tilde{\mu}_{h}}h_{k}^\dag h_{k}\cr
  &&+ \sum_{q}  \omega_{_E} \para{d_{q}^\dag d_{q}+1/2}\cr
  &&-4t\lambda_{0}\Delta_{s}^2\para{T}\sum_{q_1,q_2,q_3} h_{q_1}^\dag h_{q_2}^\dag h_{q_3} h_{-q_1-q_2-q_3}
\end{eqnarray}

where we have used the following notations:

\begin{eqnarray}
  &&\tilde{\chi}_{s}\para{T}= e^{-g^2\para{T}}\chi_{s}\para{T}\\
  && g^2\para{T} \simeq \frac{4t\gamma_{0}\Delta_{s}^2\para{T}}{ \omega_{_E}}\\
  &&4t\lambda_{0}=V=\frac{\gamma_{0}^2}{M\omega_{_E}^2}
\end{eqnarray}

in which $M$ is the ion mass. Now if we expand the energy of holons around $\vec{k}=0$ we have:

\begin{eqnarray}
&&  \epsilon_{h}\para{k}=\frac{k^2}{2m_{h}^*}-\mu^{*}_h\\
&&  m^{*}_{h}\para{T}=\frac{1}{2\tilde{\chi_{s}}\para{T}}=e^{g^{2}\para{T}}m_{h}
\end{eqnarray}

\subsection*{Oxygen isotope effect on $T_c$}

\begin{figure}
  \includegraphics[width=210pt]{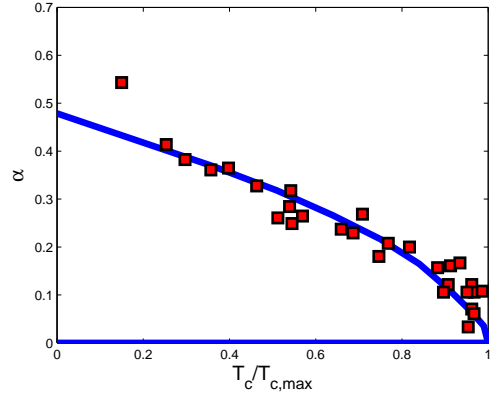}\\
  \caption{ The oxygen isotope effect on $T_c$ $(\alpha)$ vs. $T_c/T_{c,max}$. Red square are the data extracted from Fig. 1 and the blue line is our theoretical calculation}\label{fig}
\end{figure}

The oxygen isotope effect on $T_c$ is determined by the $\alpha$ isotope exponent which is defined as:

\begin{eqnarray}
  \alpha=-\frac{M_{O}}{T_{c}}\frac{d T_{c}}{dM_O}=\frac{1}{2}\frac{\omega_{_{E}}}{T_{c}}\frac{d T_{c}}{d\omega_{_{E}}}
\end{eqnarray}

In the underdoped region, $T_c$ is determined by the Bose-Einstein condensation (BEC) transition temperature of holons. If we assume holons as free 2D bose gas, then the BEC transition temperature is:

\begin{eqnarray}
  T_{c}=T_{BEC}=\frac{2\pi t p}{m^{*}_{h}\para{T_c}}
\end{eqnarray}

Therefore, we have:

\begin{eqnarray}
  \alpha=-\frac{1}{2}\frac{\omega_{_{E}}}{m^{*}_h\para{T_c}}\frac{d m^{*}_{h}\para{T_c}}{d\omega_{_{E}}}
\end{eqnarray}

Since $m^*_{h}=e^{g_{O}^2}m_{h}$ and $g_{O}^2\propto \frac{1}{\omega_{_{E}}}$ we have:

\begin{eqnarray}
  \alpha=\frac{1}{2}g_{O}^2\para{T_c}
\end{eqnarray}

From the definition of $g^2$, we have:

\begin{eqnarray}
  \alpha=\frac{2t\lambda_{0}\Delta_{s}^2\para{T_c}}{\omega_{_{O,E}}}
\end{eqnarray}

Since $\Delta_{s}\para{T_c}$ is a decreasing function of doping and at the optimal doping it becomes very small, isotope exponent dies off as we approach the optimal doping and it finally becomes negligible. On the overdoped cuprates however, pseudo-gap controls $T_c$ and as we discussed before, if $Vp^2=4t\lambda_{0}p^2\ll J$, electron phonon interaction does not affect pseudo-gap much and therefore, we do not expect isotope effect in overdoped side of the superconducting phase diagram. In our numerical calculation we have studied the phase diagram of the t-J for $J=t/3$. We have also chosen $\lambda_{0}=3\omega_{_E}/t$ ratio. We obtain a very good fitting between our theoretical calculations and the experimental data (see Fig. 3).

\subsection*{Oxygen isotope effect on the London penetration depth}

\begin{figure}
  \includegraphics[width=210pt]{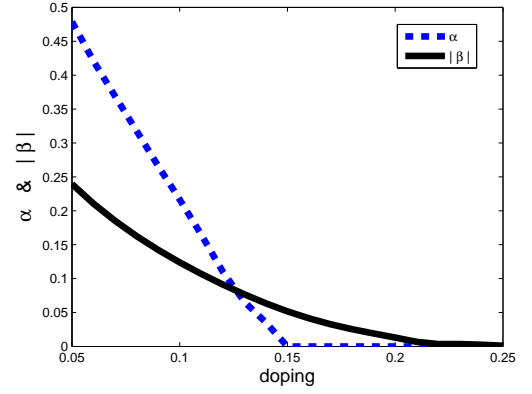}\\
  \caption{ The OIE on $T_c$, $\alpha$ (dashed blue line) and  the OIE on the London penetration depth, $\abs{\beta}$ (black line) versus doping.}\label{fig}
\end{figure}

The oxygen isotope effect on the superfluid density or equivalently on the London penetration depth, is defined as:
\begin{eqnarray}
  \beta=-\frac{M_{O}}{\lambda_{ab}}\frac{d \lambda_{ab}}{dM_O}=-\frac{1}{4}\frac{\omega_{_{E}}}{\lambda_{ab}^{-2}}\frac{d \lambda_{ab}^{-2}}{d\omega_{_{E}}}
\end{eqnarray}

\begin{figure}
  \includegraphics[width=210pt]{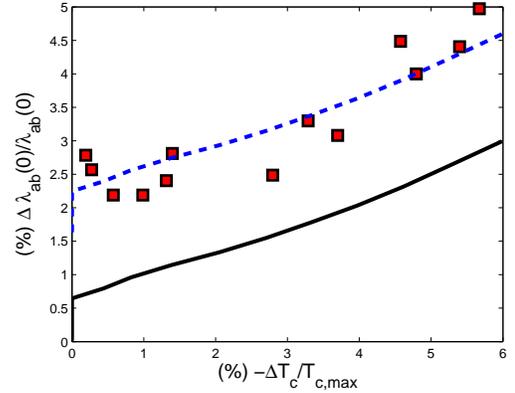}\\
  \caption{The OIE shift $\frac{\Delta \lambda_{ab}\para{0}}{\lambda_{ab}\para{0}}$  versus the OIE $-\frac{\Delta T_c}{T_c}$ (percent). Red squares are extracted data from Fig. 2. The black line is our theoretical calculation and the dashed blue line is the shifted black line. OIE on the thickness of the superconducting sheet can cause this shifting.}\label{fig}
\end{figure}

According to the Ioffe-Larkin formula$~$\cite{Ioffe_1} the physical superfluid density is related to the superfluid density of spinons and holons in the following way:

\begin{eqnarray}
  \rho_{ph}^{-1}=\rho_{h}^{-1}+\rho_{s}^{-1}
\end{eqnarray}

Since condensation fraction of holons and spinons at zero temperature are $p$ and $1-p$ respectively, and from $\rho=\frac{n_{c}}{m^*}$ we have:

\begin{eqnarray}
   \rho_{ph}^{-1}\para{0}=\frac{m^{*}_{h}}{p}+\frac{m^{*}_{s}}{1-p}
\end{eqnarray}

For small values of p, we have $\rho_{ph}\para{0}\simeq\frac{p}{m^{*}_{h}}$. On the other hand $\lambda_{ab}^{-2}\para{0}= \frac{4\pi e^2}{c^2}\rho_{ph}$. Therefore, we have:

\begin{eqnarray}
  \beta=\frac{1}{4}\frac{\omega_{_{E}}}{m^*_{h}}\frac{d  m^{*}_{h}}{d \omega_{_{E}}}=-\frac{g^2\para{0}}{4}
\end{eqnarray}

Therefore, we have:

\begin{eqnarray}
  \beta\para{T=0}=-\frac{t\lambda_{0}\Delta_{s}^2\para{0}}{\omega_{_{O,E}}}
\end{eqnarray}

For the whole superconducting region, $\Delta_{s}\para{T=0} > 0$, therefore, $\abs{\beta}$ is always nonzero though it is a decreasing function of the doping. Despite the fact that both $\alpha$ and $\beta$ depend on $\Delta_s^2$ and we may expect $\abs{\beta}=0.5\alpha$, but we should note that they depend on $\Delta_s$ at two different temperatures (see Figs. 4 and 5). From Eqs. (26) and (31), we have: $\abs{\beta}/\alpha=0.5 \para{\frac{\Delta_s\para{0}}{\Delta_s\para{T_c}}}^2\geq0.5$.

{\bf Discussion.---} Our theoretical curve for $\frac{\Delta \lambda_{ab}\para{0}}{\lambda_{ab}\para{0}}$ versus $-\frac{\Delta T_c}{T_c}$ has two important features. Firstly it has a nonzero inception since at the optimal doping, $T_c$ does not change upon isotope substitution, but $\lambda_{ab}\para{0}$ does. Secondly it has the same slope(0.42) as the empirical data but our curve is shifted down by a constant amount (see Fig. 5). In literature the cause of this shift has been discussed$~$\cite{Khasanov_2_2006}. One possible scenario is the change in the thickness of the superconducting sheet $d_s$, due to $^{16}$O/$^{18}$O substitution. Note that the 2D density of holons is related to the 3D one by: $n_{_{2D}}=n_{_{3D}} d_s$. Thus $\lambda_{ab}^{-2}\propto \frac{n_{_{3D}}d_s}{m_h^*}$ and finally we have: $\frac{\Delta \lambda_{ab}\para{0}}{\lambda_{ab}\para{0}}=0.5 \para{\frac{\Delta m^*_h}{m^*_h}-\frac{\Delta d_s}{d_s}}$. If we assume $0.5\frac{\Delta d_s}{d_s}=-1.6$ then our theoretical curve fits the experimental data very well. So our theory predicts OIE shift of the lattice spacing in the $z$ direction.

{\bf Conclusion.---} I have studied the oxygen isotope effect in cuprates. I have shown that within the spin-charge separation formalism, the unusual isotope effect on $T_c$ as well as the superfluid density can be explained by  t-J-Holstein model. It is shown that $^{16}$O/$^{18}$O substitution only affects the superfluid density and it does not affect the value of the pseudo-gap energy. This model also predicts the OIE on the thickness of the superconducting sheet.

{\bf Acknowledgement.---} I gratefully acknowledge useful discussions with X.G. Wen, T. Senthil, W. Ketterle, E. Hudson, J. Hoffman, B. Swingle, and in particular P.A. Lee for his very helpful discussions and comments. I thank R. Khasanov and H. Keller for letting me use their data and for H. Keller's clarifying remark on the experimental observations.
\bibliography{Isotope Effect}
\end{document}